\documentclass[conference]{IEEEtran}
\usepackage{amsmath}
\usepackage{amsfonts}
\usepackage{amssymb}
\usepackage{epsfig}
\usepackage{epstopdf}
\usepackage{graphicx}
\usepackage{tabularx}
\usepackage{url}
\usepackage{float}
\usepackage{stfloats}
\usepackage{multirow}
\DeclareMathOperator*{\argmax}{arg\,max}
\usepackage[section]{placeins}

\begin{document}
\title{Estimating the Number of Sources: An Efficient Maximization Approach} 

\author{
	\IEEEauthorblockN{Tara Salman\IEEEauthorrefmark{3}, Ahmed Badawy\IEEEauthorrefmark{1}\IEEEauthorrefmark{2}, Tarek M. Elfouly \IEEEauthorrefmark{3}, Amr Mohamed \IEEEauthorrefmark{3}, and Tamer Khattab \IEEEauthorrefmark{2}  }

	\IEEEauthorblockA{\IEEEauthorrefmark{1}Politecnico di Torino, DET - iXem Lab. (ahmed.badawy@polito.it)}
	\IEEEauthorblockA{\IEEEauthorrefmark{2} Qatar University, Electrical Engineering Dept. }
	\IEEEauthorblockA{\IEEEauthorrefmark{3} Qatar University, Computer Science and Engineering Dept. }
    \IEEEauthorblockA{Qatar University\\
	Doha, Qatar 2713\\
    {tara.s, badawy, tarekfouly, amrm ,tkhattab}@qu.edu.qa}
}
\IEEEoverridecommandlockouts
\IEEEpubid{\makebox[\columnwidth]{ 978-1-4799-5344-8/15/\$31.00~\copyright~2015~IEEE} \hspace{\columnsep}\makebox[\columnwidth]{ }}
\maketitle
\begin{abstract}
Estimating the number of sources received by an antenna array have been well known and investigated since the starting of array signal processing. Accurate estimation of such parameter is critical in many applications that involve prior knowledge of the number of received signals. 
Information theoretic approaches such as Akaike’s information criterion (AIC) and minimum description length (MDL) have been used extensively even though they are complex and show bad performance at some stages. In this paper, a new algorithm for estimating the number of sources is presented. This algorithm exploits the estimated eigenvalues of the auto correlation coefficient matrix rather than the auto covariance matrix, which is conventionally used, to estimate the number of sources. We propose to use either of a two simply estimated decision statistics, which are the moving increment and moving standard deviation as metric to estimate the number of sources. Then process a simple calculation of the increment or standard deviation of eigenvalues to find the number of sources at the location of the maximum value. Results showed that our proposed algorithms have a better performance in comparison to the popular and more computationally expensive AIC and MDL at low SNR values and low number of collected samples.
\end{abstract}
\section{Introduction}
\label{S:Introduction}
Estimating the number of sources in an efficient and accurate way is important to many applications that involve array signal processing. Such applications assume this parameter to be known in prior and further processing would depend on such parameter. These algorithms include: Direction of Arrival (DoA)~\cite{tuncer2009classical}, blind source and channel order separations~\cite{bell1995information}\cite{liavas1999blind}. In DoA algorithms, such as MUSIC or ESPRIT, knowing the number of sources impaired to the array is critical in eigenvalues decomposition to separate between noise and signal subspaces. DoA estimation can be involved in many further applications that include localization and tracking of objects, dedicating the signal to a desired user in wireless networks and sound and speech processing ~\cite{tuncer2009classical}.
Hence, many algorithms have been proposed to detect the number of sources that include: information theoretic criterion based~\cite{anderson1963asymptotic}~\cite{1100705}~\cite{1164557}, eigenvector-based~\cite{1172651},and threshold based estimations~\cite{1311595}.

Information theoretic approaches such as Akaike’s information criterion (AIC)~\cite{1100705} and minimum description length (MDL)~\cite{1164557} are the most widely used methods for number of sources estimation. Those methods are criterion based estimation algorithms that are mostly computationally complex and have bad performance with low number of samples and low SNR.
Complexity problem that is found in both methods is due to the minimization of criterion to search for minimum AIC or MDL values beside the eigenvalue decomposition (EVD) operation on the auto covariance matrix of the observed data. The poor performance problem is due to the incorrect estimation of auto covariance matrix at low SNR especially with low number of samples. This results in no clear difference between eigenvalues that are needed for number of sources estimation. Beside all that, such methods assume the noise to be sparse-like uncorrelated from the signal and hence fail in practical scenarios such as underwater~\cite{kotanchek1996subspace} and indoor offices~\cite{jiang2002path}. As a result, a lot of research tried to solve these problems by modifying the traditional algorithms or proposing another way for estimation.

Some works tried to reduce the complexity by not going through EVD and solving the problem utilizing Multi-Stage Wiener Filter (MSWF),~\cite{an2008new}, however, applications such as DoA would mostly involve EVD so going through that wouldn't add complexity. 
Another work in \cite{gu2007detection} presented a threshold based estimation algorithm that is based on peak to average ratio (PAR) characteristics. The algorithm calculates the PAR values of the received data and get the differences between adjacent ones which is compared to a threshold. If the difference exceeds the threshold, then the number of sources is detected by the location at that point. The threshold here is set based on the gradient of the PAR values, the number of array elements and the minimum PAR value. Results showed a better performance than AIC and MDL under low SNR conditions, however the threshold need to be adjusted and the probability of detection is still effected drastically by the number of samples.

In this paper, it was noted that most information theoretic approaches are computationally complex while threshold based approaches need a reconfigurable threshold with different parameters. As well, almost all previous work did consider the auto covariance matrix eigenvalues without considering the auto correlation coefficient matrix eigenvalues which can result in a much simpler detection approaches. Hence, this paper proposes a simple estimation algorithm that uses the auto correlation coefficient matrix to estimate eigenvalues and estimate the number of sources by looking for the maximum difference or moving standard deviation between eigenvalues. The moving standard deviation in here is the difference between two consecutive biased standard deviation of two eigenvalues only. The algorithm was compared, in term of error rate, to information theoretic approaches with different scenarios and setups. 

The rest of the paper is organized as follows, section~\ref{sec:System model} will present the system model while section~\ref{sec:prob} presents how the eigenvalues are calculated. Section~\ref{sec:existing} presents some existing techniques and section~\ref{proposed2} will present the proposed algorithm. Section~\ref{sec:Simulation} will present simulation results and comparisons and finally, conclusion and future work will be presented in section~\ref{sec:Conclusion}.
\section{System Model }
\label{sec:System model}
In our system model, we assume that the receiver is equipped with $M$-sensor uniform antenna array. Considering $K$ signals are impinging on the receiver's array, the received signal at an instant of time $t$ can be expressed as:
\begin{equation}
\mathbf{y(t)=\sum_{k=1}^{K}{a(\theta_k)*s_k(t)}+w(t)}, \label{eqn0}
\end{equation}
\noindent where $\mathbf{a(\theta_k)}$ is the steering vector for the a signal arriving at azimuth angle $\theta_k$, $\mathbf{s_k(t)}$ is the impinging signal from the $k^{th}$ source at time $t$, and $\mathbf{w(t)}$ is the additive white Gaussian noise (AWGN). In the matrix notation, (\ref{eqn0}) can be represented as:
\begin{align}
\mathbf{Y=AS+W},
\end{align}
\noindent where $Y\in \mathbb{C} ^{M \times N}$, $\mathbf{A}\in \mathbb{C} ^{M \times K}$,$ \mathbf{S}\in \mathbb{C} ^{K \times N}$, $ \mathbf{W}\in \mathbb{C} ^{M \times N}$, with $N$ being the total number of collected samples and $\mathbb{C}$ is the set of complex numbers. The matrix of steering vectors is:
\begin{equation}
\mathbf{A=[a(\theta_1),a(\theta_2)...a(\theta_k)]}.
\end{equation}
The steering vector $\mathbf{a(\theta_k)}$ for a uniform circular array (UCA) can be represented as:
\begin{equation}
\label{circular}
 \mathbf{a(\theta_k)}=e^{\left(2 \pi/\eta r \left(cos\left( \theta-\gamma \right)\right)\right)},
\end{equation}
\noindent with waveform $\eta$, radius r and $\gamma$ is $360/N*[0:N-1]$.

The auto covariance matrix $R$ of the received data can be expressed as:
\begin{eqnarray}
\mathbf{R_{YY}}&=& \mathbb{E}\left[\mathbf{YY^H}\right]\\ \nonumber
 &=&\mathbf{AR_{SS}A^H}+\mathbf{R_{WW}} \label{eqncov}
\end{eqnarray}
\noindent where, $\mathbb{E}[.]$ denotes the expectation operation, $H$ denotes the Hermitian operation, $\mathbf{R_{SS}}$ is the auto covariance matrix of the impinging signal, $\mathbf{R_{WW}}=\sigma^2 \mathbf{I}$ is the auto covariance matrix of the receivers AWGN with $\sigma^2$ is the noise variance and $\mathbf{I}$ is $M\times M$ unitary matrix. It is worth noting that the auto covariance matrix of the impinging signal $\mathbf{R_{SS}}$ is assumed to be a full rank matrix. This implies that its columns are linearly independent or in other words, the impinging signals are not correlated. Consequently, if the impinging signals are correlated, $\mathbf{R_{SS}}$ will be rank deficient.

\section{Problem Formulation} \label{sec:prob}
The auto covariance matrix of the received signal from the $M$ antenna array is typically estimated when estimating the DoA \cite{BARTLETT, Capon}. For subspace based techniques such as MUltiple SIgnal Classification (MUSIC) \cite{Phd1981}, which is widely used and known for its superb performance particularly at low SNR levels, the EVD is applied on $\mathbf{R_{YY}}$ as a step to estimate the DoA. In other words, estimating the $\mathbf{R_{YY}}$ and its EVD is a conventional step in most of the DoA estimation algorithms. Applying EVD on $\mathbf{R_{YY}}$ leads to:

\begin{eqnarray} \label{eqn01}
\mathbf{R_{YY}}&=&\mathbf{U_Y\Lambda_Y U_Y^H}\\ \nonumber
&=&\mathbf{U_S\Lambda_S U_Y^H + U_W\Lambda_W U_W^H},
\end{eqnarray}
\noindent where $\mathbf{U_S}$ and $\mathbf{U_W}$ are signal and noise subspaces unitary matrices, and $\mathbf{\Lambda_S}$ and $\mathbf{\Lambda_W}$ are diagonal matrices of the eigenvalues of the signal and noise, respectively. (\ref{eqn01}) can be expressed as:
\begin{align}
\mathbf{U_Y \Lambda_Y U_Y^H} = diag\left(\lambda_1 , \lambda_2 , ... \lambda_K, 0 , ..., 0 \right) + \sigma^2 \mathbf{I}. \label{eqn1}
\end{align}


The eigenvalues $(\lambda_1 , \lambda_2 , ... \lambda_M )$  with their corresponding eigenvectors $(e_1 , e_2,...e_M )$ define the signal and noise subspace as $\mathbf{U_S}=[e_1,...,e_K ]$ and $\mathbf{U_W}=[e_{K+1},....,e_M]$ respectively. The problem is then estimating the value of $K$, i.e., the number of impinging signals,  given the estimated $(\lambda_1, \lambda_2,... \lambda_M) $.
\section{Existing Techniques}\label{sec:existing}
AIC and MDL are the most widely used algorithms for number of sources estimation. They are order determination information theoretic models that use the eigenvalues of the sample auto covariance to determine how many smallest eigenvalues are approximately equal. Those eigenvalues would lie in the noise subspace while others would lie in signal subspace. Both algorithms consist of minimizing a criterion of log likehood over the number of signals that are detectable. In here, the derivation of those criterion will not be stated, however the details of both of them can be found in~\cite{1100705}. When ordering the eigenvalues in a descending order, i.e., $\lambda_1 \geq \lambda_2 \geq ... \lambda_M$, AIC criterion can be expressed as:
\begin{equation}\begin{split}
K_{AIC}& = argmin_k \bigg(-2 \log\left(\frac{\Pi_{i=k+1}^M\lambda_i^{\frac{1}{M-k}}}{\frac{1}{M-k}\sum_{i=k+1}^M\lambda_i}\right)^{(M-k)N}+ \\& 2k(2M-k)\bigg) \label{eqnAIC}
\end{split}
\end{equation}
while MDL criterion can be expressed as:
\begin{equation}
\begin{split}
K_{MDL}=& argmin_k \bigg(- \log\left(\frac{\Pi_{i=k+1}^M\lambda_i^{\frac{1}{M-k}}}{\frac{1}{M-k}\sum_{i=k+1}^M\lambda_i}\right)^{(M-k)N} \\ \label{eqnMDL}
& + \frac{1}{2}k(2M-k)\log(N)\bigg)
\end{split}
\end{equation}
where $k$ is the index of the eigenvalues. For the rest of the paper, we will use AIC and MDL as references to compare the performance of our proposed algorithms.
Another approach for estimating the number of sources is based on setting a threshold for the eigenvalues increment \cite{hu1999detecting}. It was noted that the eigenvalues of the noise subspace are close to each other and the difference between them doesn't exceed a certain threshold. Hence, the increment in the eigenvalue is compared with a threshold to estimate the number of sources. Their estimated threshold ($\gamma_{inc}$) is given by:
\begin{equation}
 \gamma_{inc}=\rho(M,N) \frac{P_s}{(1+\sqrt{P_s/\lambda_M})^2}
\end{equation}
where $P_s$ is the estimated signal power, $\lambda_M$ is the eigenvalue with index M, i.e. last eigenvalue. $\rho$ in here is a coefficient that is found through extensive computer simulation for each two particular $M$ and $N$. In other words, each time either $N$ or $M$ or both of them change, a comprehensive simulation has been run beforehand to find the best $\rho$ value.

AIC and MDL are more computationally expensive than the eigenvalues increment threshold based approach given that it will be needed to solve the minimization problem given in (\ref{eqnAIC}) and (\ref{eqnMDL}) each time an estimation of the number of sources in needed. On the other hand, the eigenvalues increment threshold based approach requires an extensive iterations a prior to adjust $\rho$ accordingly. In addition to that $\rho$ depends on several parameters such as $N$, $M$ and SNR making its adjustment a tedious process.

\section {Proposed Algorithm} \label{proposed2}
In our proposed algorithm, we exploit the auto correlation coefficient matrix rather than the auto covariance matrix to estimate the number of impinging sources. To define the auto correlation coefficient matrix, we first redefine the auto covariance matrix in (\ref{eqncov}) as:
\begin{eqnarray}
\mathbf{V_{YY}}&=& \mathbb{E}\left[\left(\mathbf{Y}-\mathbf{\mu_Y}\right)\left(\mathbf{Y}-\mathbf{\mu_Y}\right)^H\right]\\ \nonumber
 &=&\mathbf{AR_{SS}A^H}+\mathbf{R_{WW}}-\mathbf{\mu_Y \mu_Y^H}
\end{eqnarray}
\noindent where $\mathbf{\mu_Y}=\mathbb{E}[\mathbf{Y}]$. The elements in the diagonal of $\mathbf{V_{YY}}$ are the variances of $\mathbf{Y}$. The auto correlation coefficient matrix $\mathbf{C_{YY}}$ is then given by:

\begin{align}
\mathbf{C_{YY}=\left(diag(V_{YY})\right)^{-\frac{1}{2}}V_{YY}\left(diag(V_{YY})\right)^{-\frac{1}{2}}}.
\end{align}
\noindent We then apply the EVD on $\mathbf{C_{YY}}$ which leads to:

\begin{align}
\mathbf{C_{YY}=U_C \Lambda_C U_C^H},
\end{align}
\begin{equation}
\begin{split}
\mathbf{U_C \Lambda_C U_C^H} =& diag\left(\lambda_1^C , \lambda_2^C , ... \lambda_K^C, 0 , ..., 0 \right)\\
& + \left(diag(\mathbf{V_{YY}})\right)^{-\frac{1}{2}} \left(\sigma^2 \mathbf{I}-\mathbf{\mu_Y \mu_Y^H}\right)\\
&\left(diag\mathbf{(V_{YY}})\right)^{-\frac{1}{2}}. \label{eqncorrEVD}
\end{split}
\end{equation}

\noindent The eigenvalues $(\lambda_1^C , \lambda_2^C , ... \lambda_M^C )$  with their corresponding eigenvectors $(e_1^C , e_2^C,...e_M^C )$ define the signal and noise subspace as $\mathbf{U_S}=[e_1^C,...,e_K^C ]$ and $\mathbf{U_W}=[e_{K+1}^C,....,e_M^C]$ respectively. As well, the problem is then estimating the value of $K$ given the estimated $(\lambda_1^C, \lambda_2^C,... \lambda_M^C)$. We first arrange the eigenvalues in an ascending order, rather than a descending order as in the case of AIC and MDL. Hence, eigenvalues are arranged from the beginning as $(\lambda_1^C, \lambda_2^C,... \lambda_M^C) $ where $\lambda_1^C \le \lambda_2^C \le... \lambda_M^C $ and  $(\lambda_1^C, \lambda_2^C,... \lambda_{M-K-1}^C)$ would lay in the noise subspace while $(\lambda_{M-K}^C... \lambda_M^C)$ are in signal subspace.

It can be inferred from (\ref{eqncorrEVD}) and (\ref{eqn1}) that since the eigenvalues of the signal subspace contain both signal and noise power, the values of sources' signal eigenvalues are expected to be higher than noise eigenvalues at moderate and high SNR values. At the same time the noise eigenvalues are expected to be comparable to one another. The main contribution of using EVD of the auto correlation coefficient matrix in (\ref{eqncorrEVD}) rather than EVD of the auto covariance matrix in (\ref{eqn1}) is that the difference between signal eigenvalues and the noise eigenvalues is more accentuated, which leads to an easier and more efficient estimation of the number of sources, particularly at low SNR values. Moreover, the mathematical operation applied to estimate a decision statistic, which is then used to decide on the number of sources, can be as simple as our proposed moving increment or moving standard deviation rather than the complicated decision statistic for the AIC and MDL given in (\ref{eqnAIC}) and (\ref{eqnMDL}).

\begin{figure}[!t]
\centering
\includegraphics[width=0.46\textwidth]{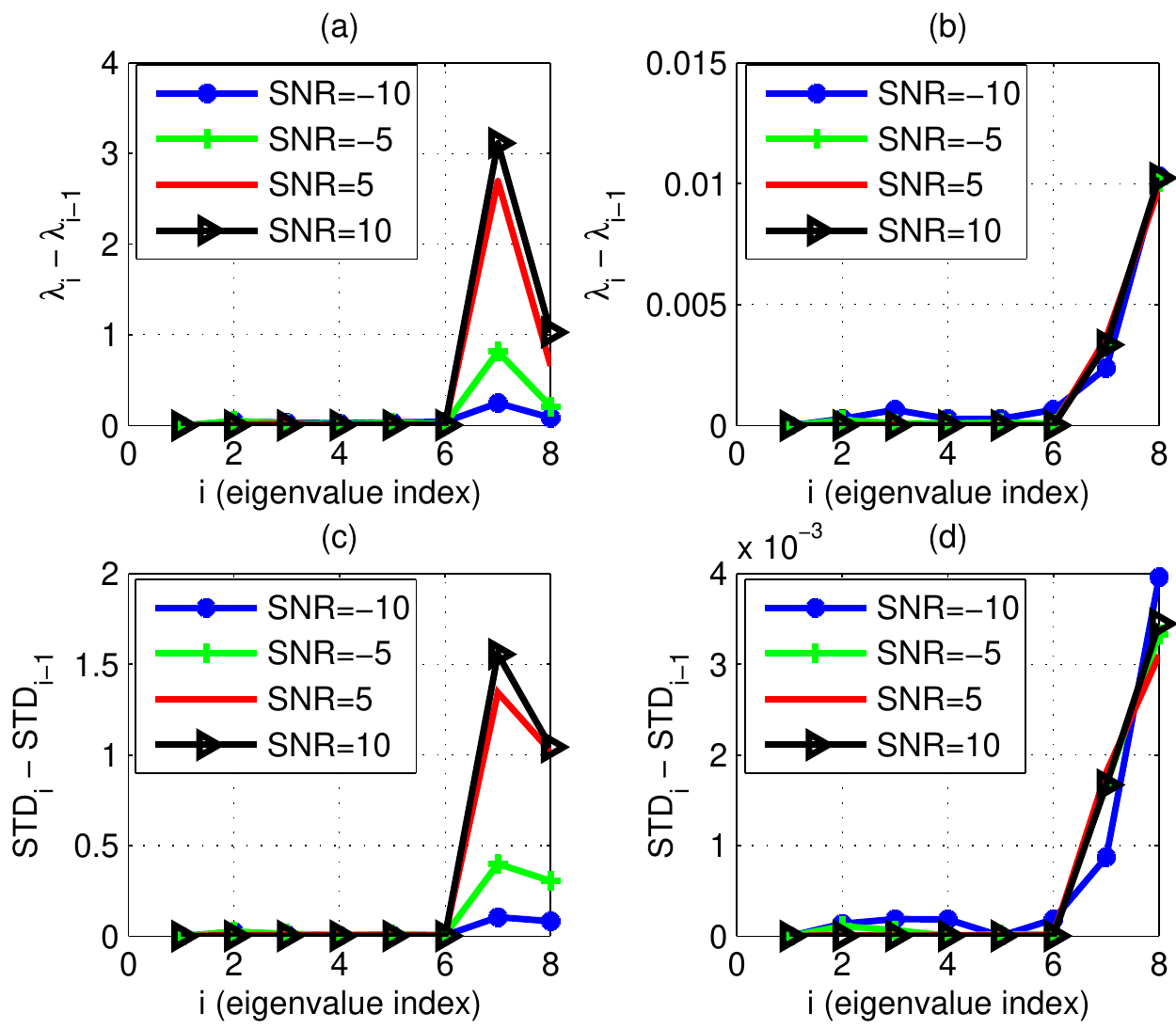}
\caption{Change in Eigenvalues decision statistics with different SNR (a) Moving Increment of auto correlation coefficient matrix (b) Moving Increment of auto covariance matrix (c) Moving Standard Deviation of auto correlation coefficient matrix (d) Moving Standard Deviation of auto covariance matrix }
\label{F:eigenvalues_SNR}
\end{figure}

\begin{figure}[!t]
\centering
\includegraphics[width=0.46\textwidth]{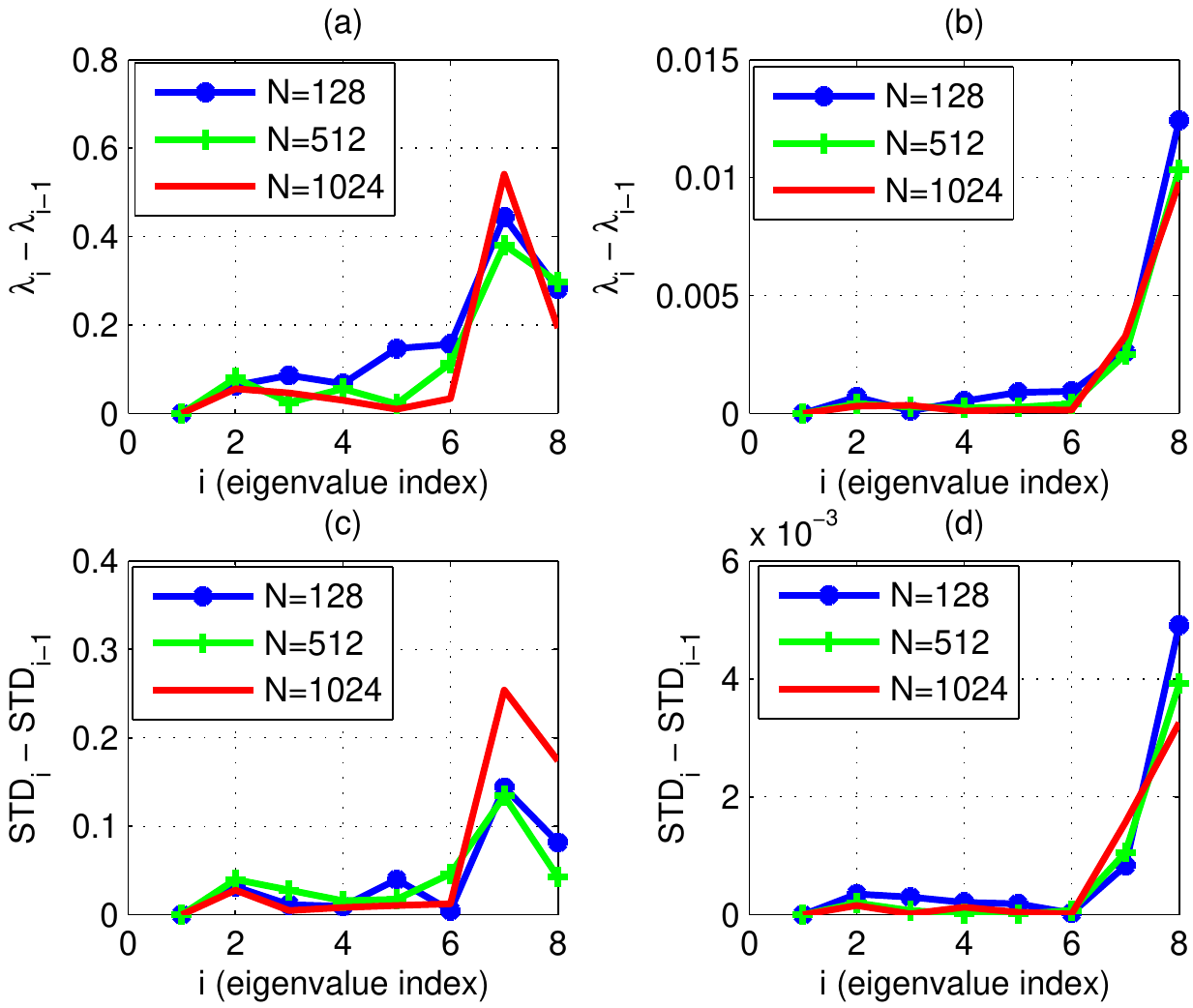}
\caption{Change in Eigenvalues decision statistics with different number of samples (a) Moving Increment of auto correlation coefficient matrix (b) Moving Increment of auto covariance matrix (c) Moving Standard Deviation of auto correlation coefficient matrix (d) Moving Standard Deviation of auto covariance matrix}
\label{F:eigenvalues_L}
\end{figure}

To illustrate the advantage of using  EVD of the auto correlation coefficient matrix, as we propose, versus EVD of the auto covariance matrix, which is conventionally used in most of the existing techniques, we plot the moving increment and moving standard deviation of the estimated eigenvalues of the auto correlation coefficient matrix versus the auto covariance matrix for different SNR values in Fig.\ref{F:eigenvalues_SNR} and different number of collected samples in Fig.\ref{F:eigenvalues_L}. The simulation parameters for the first figure are: 8 elements antenna array, 2 impairing signal, 1024 samples and different SNR values. The simulation parameters for the other figure are the same except that the SNR is kept fixed at -7 dB and the number of samples changed from 128 to 1024. From these figures, one can see that in the case of using the eigenvalues of the auto correlation coefficient matrix, the jump in the decision statistic when first moving from the noise subspace to the signal subspace is always the highest. The decision statistics then starts to decrease, in other words, the highest increment in the decision statistic always happens when moving from noise subspace to signal subspace. On the contrary, when using the same two decision statistics with the eigenvalues of the auto covariance matrix, the first jump between the noise and signal subspaces is not necessarily the largest. In addition to that the decision statistics in the signal subspace increase monotonically. This implies that when using the decision statistics of the auto correlation coefficient matrix, the problem is transformed into a simple maximization problem, where the index at which the highest jump occurs is searched for. While for the case of using the decision statistic of the auto covariance matrix, the decision statistics should be compared to a threshold to decide on the number of sources. As stated in \cite{hu1999detecting}, which uses covariance eigenvalues in their algorithms, estimating the threshold is a tedious process that requires an extensive simulation and iterations to estimate the appropriate threshold for each particular set of parameters.

\subsection{Moving Increment of the auto correlation coefficient matrix Eigenvalues}
Our first proposed decision statistic ($\delta$) used as a metric to decide on the number of sources is the moving increment of the estimated eigenvalues of the auto correlation coefficient matrix. The moving increment is estimated as the difference between each two consecutive eigenvalues:
\begin{equation}
\label{Diff}
\delta_{i}={\lambda_i^C-\lambda_{i-1}^C} \hspace{0.2in} \text{for} \hspace{0.2in} i=2,3,.., M.
\end{equation}
The highest increment would then imply the shift between noise eigenvalues to signal eigenvalues. The index at which this shift happens can be estimated as:
\[
 j=\argmax_i \delta_{i}.
\]
\noindent In this case, the number of sources can be given by $K=M-j+1$.
\subsection{Moving STD of the auto correlation coefficient matrix Eigenvalues}
Our second proposed decision statistic ($\alpha$) used as a metric to decide on the number of sources is the moving standard deviation of the estimated eigenvalues of the auto correlation coefficient matrix. The biased sample standard deviation in general is a measure of variance or difference of the sample from the mean, it can be calculated by:
\begin{equation}
s_M= \sqrt{\frac{1}{M-1} \sum_{i=1}^{M}{\left( x_i-u \right)}^2},
\end{equation}
where $u$ is the mean and $M$ is the size of the sample or, in our case, the size of the eigenvalues involved in standard deviation calculation.

\noindent Now, finding the biased standard deviation of two eigenvalues, can be done by:
\begin{equation}
\label{STD}
STD(i)=\sqrt{{(\lambda_i^C-u)^2+(\lambda_{i-1}^C -u)^2}},
\end{equation}
\noindent where $u$ is the mean of the two eigenvalues involved which is:
\begin{equation}
\label{mean}
u=\frac{\lambda_i^C+\lambda_{i-1}^C}{2}.
\end{equation}
\noindent We define our second decision statistic, which is the moving STD ($\alpha$) as the difference between two consecutive STDs:
\begin{eqnarray}
\alpha_i &=& STD(i)-STD(i-1) \nonumber \hspace{.25 in}\text{for} \hspace{.05 in}i=3, 4, ... , M \\
&=&\left(\frac{(\lambda_i^C-\lambda_{i-1}^C)-(\lambda_{i-1}^C-\lambda_{i-2}^C)}{\sqrt{2}}\right).
\end{eqnarray}
\noindent Similarly, as in the case of using the moving increment, the highest index at which the shift between the noise eigenvalues and the signal eigenvalues can be estimated as: 
\[
 j=\argmax_i \alpha_i
\]
\noindent Consequently, the number of sources can be given by $K=M-j+1$.

\section{SIMULATION RESULTS }
\label{sec:Simulation}
Simulation results were carried in different scenarios to test algorithm's performance with different cases that include different SNR values, different number of samples, different number of impairing signals, and different array configuration. Performance metric used for comparison was the percentage error rate, which can be expressed as:
\begin{equation}
error~rate = \left(1- \frac{number~of~successes}{ number~of~runs}\right) \times 100
\end{equation}

Except for the last simulation, the array that was used was a uniform circular array with 8 elements. The original signal was a QPSK signal and the noise added was a white Gaussian noise with the different SNR values. The number of runs iterations is 10000.

\subsection {Algorithms Performance at Various SNR}
The first simulation was done to test AIC, MDL and the proposed algorithms performance with different SNR values. SNR values ranged from -20 to 15 dB, the number of samples was fixed to 1024 and the actual number of sources was 2.
\begin{figure}[!t]
\centering
\includegraphics[width=0.4\textwidth]{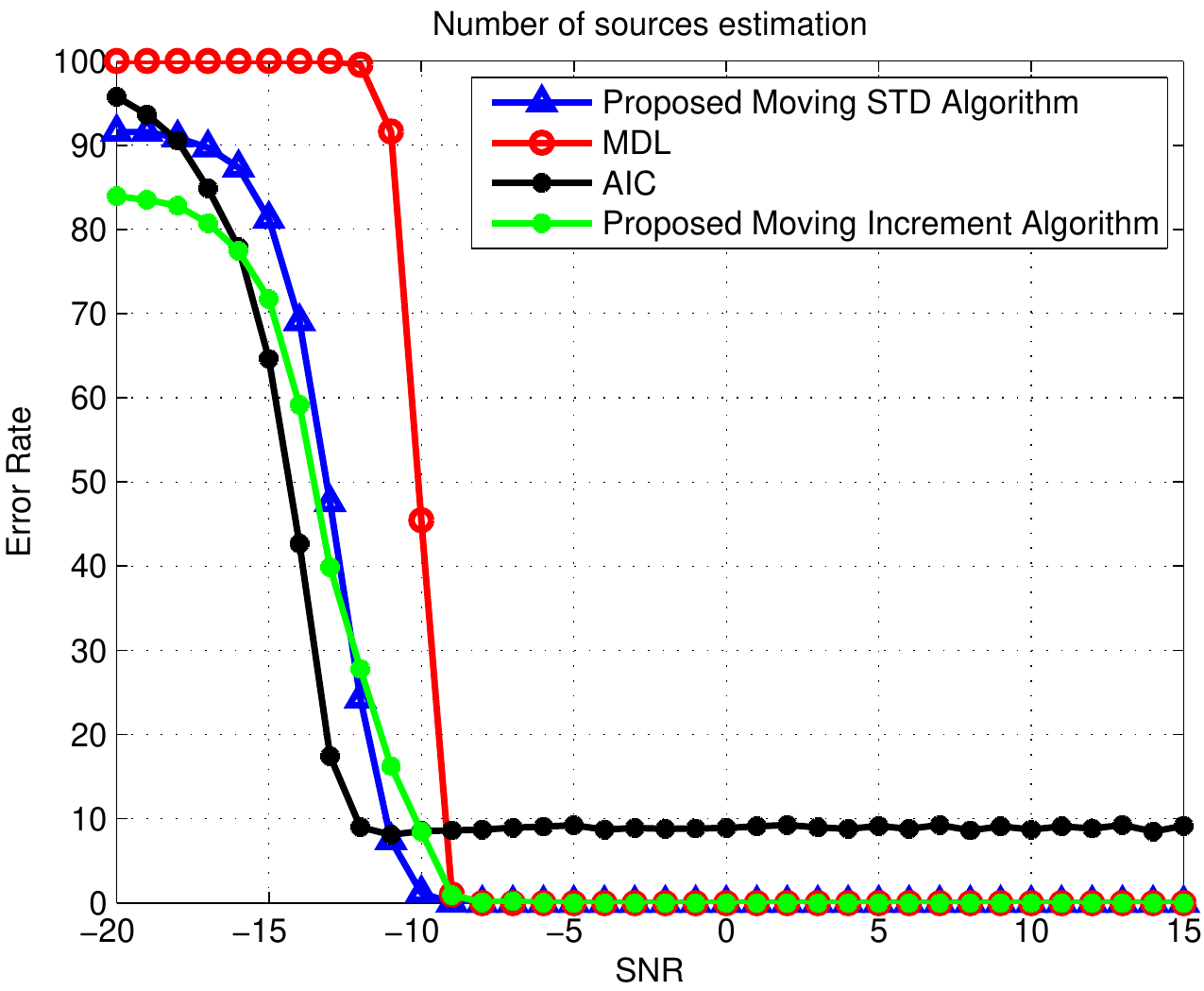}
\caption{Effect Of Different SNR at 1024 sample}
\label{F:L1024}
\end{figure}

As shown in Fig.~\ref{F:L1024}, the proposed algorithms behaved better than MDL in low SNR values and better than AIC in high SNR values. For less than -10dB, the performance of the proposed algorithms had a comparable performance to AIC and better than MDL. However, the estimation was inaccurate for all algorithms for less than -12 dB SNR. The reason for the proposed algorithms inaccurate estimation at this stage is due to the inconsistent change in eigenvalues that is resulted from high noise and hence the standard deviation or increment changes randomly and the detection can happen at different stages. The reason why MDL behaves badly is the underestimation of the number of sources which was detected to be 1 as well. After -10 dB SNR, the performance of MDL and the proposed algorithms came to be the same with minimum error rate that is almost 0 while AIC kept its error rate of about 10. The reason why AIC is not giving lower error rate is the overestimation of the number of sources which happen with relatively high SNR values. This overestimation is probably due to AIC added penalty term as was proven by \cite{995060}. 

\subsection {Algorithms Performance at Various Number of Samples}
One of the important parameters to consider in any algorithm design is the number of samples needed by the algorithm to estimate correctly. This is important for algorithm practical implementation as the number of samples needed to be minimized in such scenarios. Hence, MDL, AIC and the proposed algorithms were tested against different number of samples at SNR value of -5 dB with 2 impairing signals.

\begin{figure}[!t]
\centering
\includegraphics[width=0.4\textwidth]{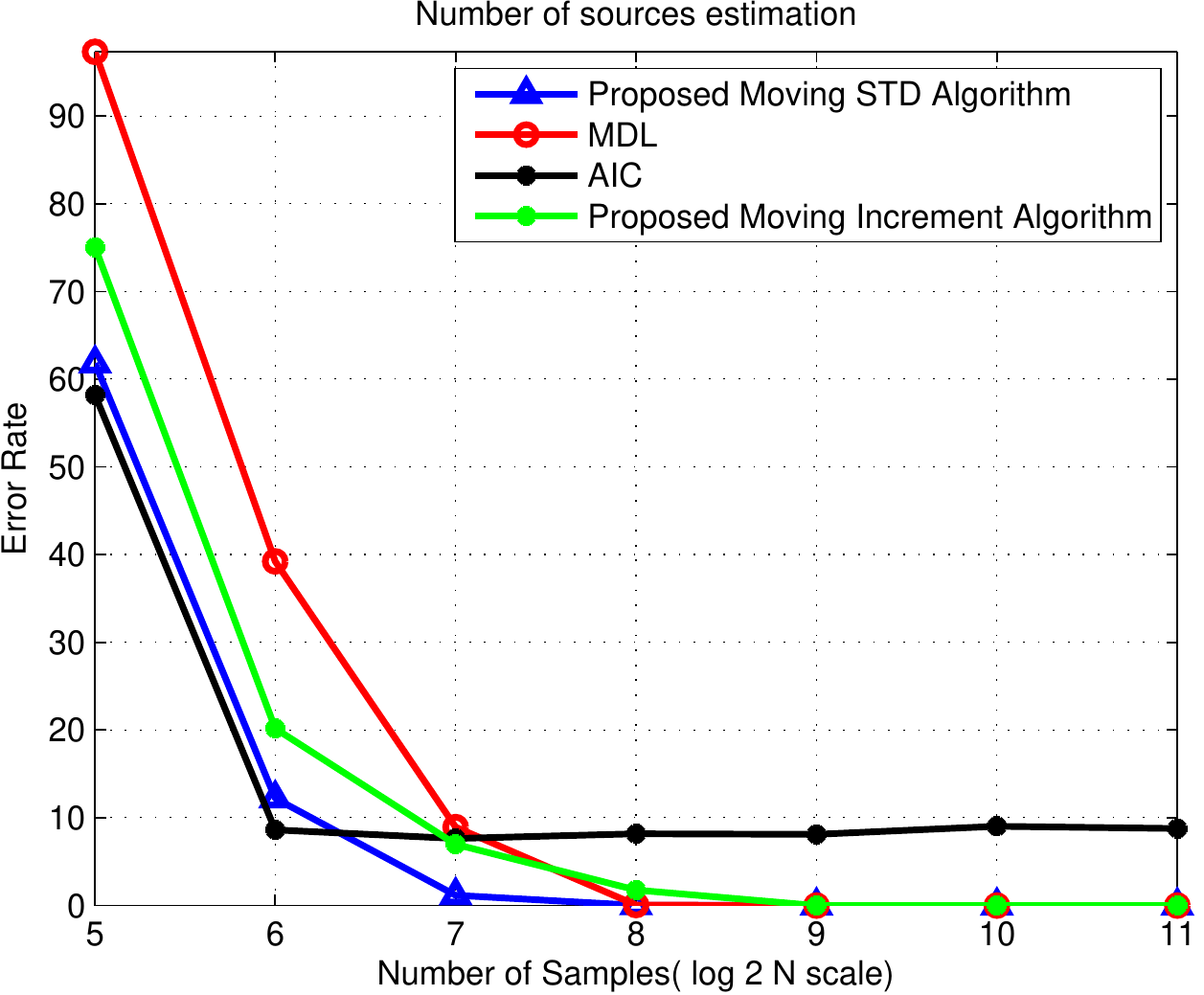}
\caption{Effect Of Different Number of samples at SNR=-5}
\label{F:SNR-5}
\end{figure}

As can be seen in Fig.\ref{F:SNR-5}, the proposed algorithms did have a better performance than MDL and similar performance to AIC for low number of samples. MDL algorithm underestimated the number of sources with low number of samples as eigenvalues were not well distrusted in a way that can be detected by the algorithm criterion and the added penalty term. MDL and the proposed algorithms did have the same performance for more than 256 samples which was almost 0 error rate. AIC, on the other hand, overestimated the number of samples and hence had its 10\% error rate, which was found in almost all test cases that were conducted in this paper.

\subsection {Algorithm Performance with Different Number of Impairing Sources }
Different algorithms might have different sensitivities in terms of the number of sources they can estimate. Hence, algorithms were tested against a different number of impairing sources at SNR value of -5 dB and the number of samples equal to 1024 samples.

\begin{figure}[!t]
\centering
\includegraphics[width=0.4\textwidth]{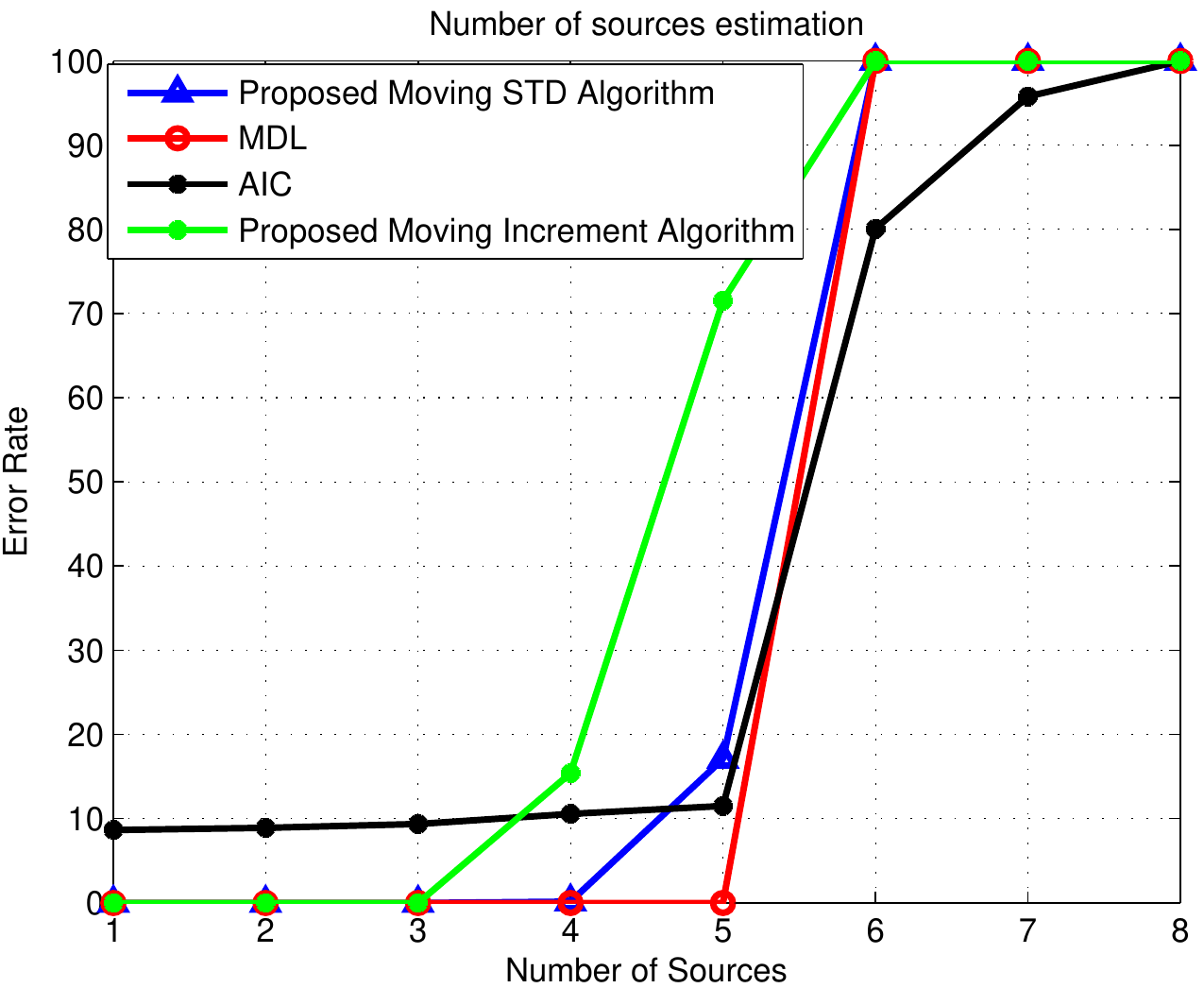}
\caption{Effect Of Different Number of Impairing Signals at SNR=-5, L =1024 with 8 elements UCA array}
\label{F:Signals2}
\end{figure}

As can be seen in Fig.~\ref{F:Signals2}, AIC outperformed all other algorithms in the maximum number of sources it can estimate. MDL and moving STD algorithm could estimate up to 5 sources with less than 20\% error and fail to estimate more while AIC could estimate 6 and 7 but with high error rate. The reason behind this drawback goes back to separation between the DoA angles was not enough to estimate the number of sources correctly at this point. In general, moving STD algorithm could estimate up to 6 signals with 8 elements array and couldn't estimate more no matter what the separation or SNR value was. Moving Increment algorithm could estimate up to 5 sources with this configuration. However, such performance drawback can be safely negligible due to the fact that receiving 6 sources at the same time is almost impossible in practical wireless scenarios. Besides, even if the number of sources was estimated correctly, further applications that use such estimation, such as DoA, won't be able to estimate more than 5 sources and hence the difference in the performance won't effect.

\subsection {Algorithm performance with Different Array Elements}
This simulation examines the effect of increasing the number of elements that construct the array. The test was done on the SNR value of -5, 100 samples and 2 signals were impaired to the array.
\begin{figure}[!t]
\centering
\includegraphics[width=0.4\textwidth]{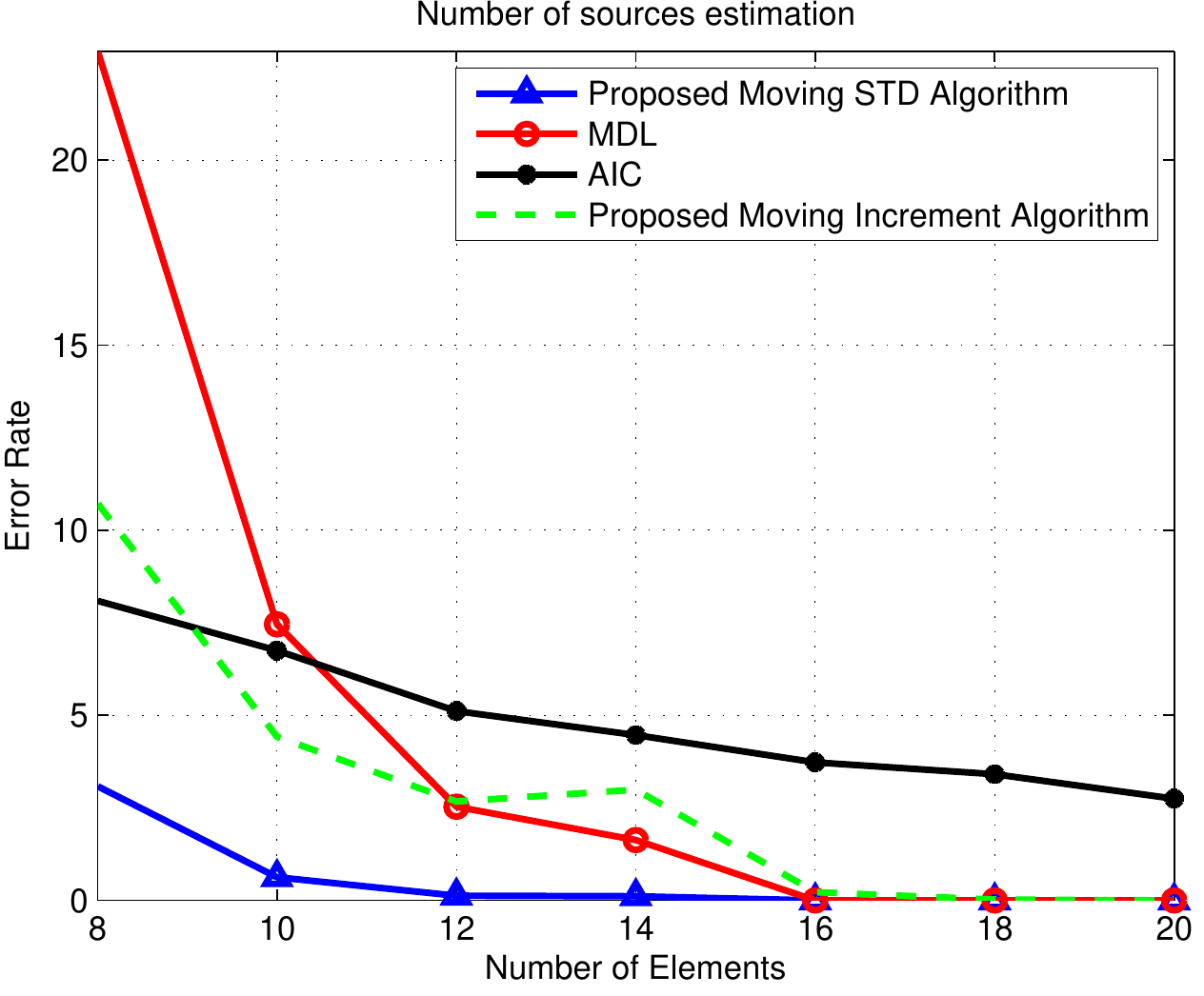}
\caption{Effect Of increasing the number of elements that construct the array at L =100}
\label{F:Elements2}
\end{figure}

As shown in Fig.~\ref{F:Elements2}, when the number of elements increases the error rate will decrease until it reach almost 0 error rate when the number of elements is 8 for the moving STD based algorithm and 12 for MDL and moving increment based algorithm.  Before 16 elements array moving STD algorithm showed the best performance compared to three simulated ones which can be resulted back to low number of samples that caused MDL and moving increment to perform worst than others at that stage. After 16 elements both MDL and the proposed algorithms showed an error rate of almost 0\%. 
\section{Conclusion}
\label{sec:Conclusion}
This paper presented a new algorithm for number of sources estimation based on eigenvalues decomposition. The algorithm depended on the auto correlation coefficient matrix instead of auto covariance matrix to get the eigenvalues and find the number of samples at the maximum moving increment or moving standard deviation. In general, moving STD behaved better than moving increment and had more stable performance that is comparable to AIC and MDL. Results showed a better performance than MDL at low SNR values and better than AIC at high SNR. In addition to that, the proposed algorithm was much simpler than the information theoretic approaches as it depends on simple maximizing problem only.

\section*{Acknowledgment}
This publication was made possible by the support of the NPRP grant 5-559-2-227
from the Qatar National Research Fund (QNRF). The statements made herein are solely the
responsibility of the authors.
\bibliographystyle{IEEEtran}
\bibliography{paper}
\end{document}